\documentclass[conference]{IEEEtran}
\IEEEoverridecommandlockouts
\usepackage{cite}
\usepackage{amsmath,amssymb,amsfonts}
\usepackage{siunitx}
\usepackage{algorithm,algorithmic}
\usepackage{graphicx}
\usepackage{textcomp}
\usepackage{xcolor}
\usepackage{hyperref}
\usepackage{tabularx, booktabs}
\usepackage{marvosym}
\usepackage{amssymb}
\usepackage{cleveref}
\usepackage{pifont} 

\def\BibTeX{{\rm B\kern-.05em{\sc i\kern-.025em b}\kern-.08em
    T\kern-.1667em\lower.7ex\hbox{E}\kern-.125emX}}
\begin{document}

\title{DNN-based Digital Twin Framework of a DC-DC Buck Converter using Spider Monkey Optimization Algorithm\\
}

\author{
	\IEEEauthorblockN{Tahmin Mahmud\IEEEauthorrefmark{1} and Euzeli Cipriano Dos Santos Jr.}
	\IEEEauthorblockA{
		Elmore Family School of Electrical and Computer Engineering \\
		Purdue University \\
		Indianapolis, IN, USA \\
		Email: \{mahmud13, edossant\}@purdue.edu
	}
}

\maketitle

\begin{abstract} Component ageing is a critical concern in power electronic converter systems (PECSs). It directly impacts the reliability, performance, and operational lifespan of converters used across diverse applications, including electric vehicles (EVs), renewable energy systems (RESs) and industrial automation. Therefore, understanding and monitoring component ageing is crucial for developing robust converters and achieving long-term system reliability. This paper proposes a data-driven digital twin (DT) framework for DC-DC buck converters, integrating deep neural network (DNN) with the spider monkey optimization (SMO) algorithm to monitor and predict component degradation. Utilizing a low-power prototype testbed along with empirical and synthetic datasets, the SMO+DNN approach achieves the global optimum in 95\% of trials, requires 33\% fewer iterations, and results in 80\% fewer parameter constraint violations compared to traditional methods. The DNN model achieves $R^2$ scores above 0.998 for all key degradation parameters and accurately forecasts time to failure. SMO-tuned degradation profile improves the converter's performance by reducing voltage ripple by 20--25\% and inductor current ripple by 15--20\%.
\end{abstract}

\begin{IEEEkeywords}
 buck converter, digital twin, spider monkey optimization (SMO), deep neural network (DNN), reliability analysis.
\end{IEEEkeywords}
\vspace{-0.5cm}
\section{Introduction}
Buck converters are essential to North America’s push for electrification, clean energy, and digital infrastructure. This topology supports key technologies e.g., point-of-load regulation \cite{b6}, on-board chargers \cite{b7}, auxiliary supplies in EVs \cite{b8}, and power conditioning units in battery storage systems \cite{b9}. The emergence of wide-bandgap semiconductors like GaN and SiC has further expanded research opportunities to improve thermal efficiency and power density \cite{b10}. Moreover, federal investments in grid modernization and electric transportation continue to drive demand for robust, compact, and intelligent power converters \cite{b11}.

As buck converters find broader use in fault diagnosis, maintenance, and condition monitoring, ensuring their long-term reliability is increasingly critical. This brings attention to the issue of component ageing, a gradual degradation of electrical and mechanical properties in the passive components due to prolonged exposure to environmental and operational stresses such as high temperatures, voltage transients, thermal cycling, vibration, and humidity \cite{b12}. These conditions trigger failure mechanisms including solder joint fatigue, dielectric breakdown in capacitors, increased ON-state resistance ($r_{ds\text{-}ON}$), and degradation of magnetic core materials \cite{b13,b14}. Thus, understanding ageing effects is crucial for extending buck converter's lifespan and ensuring stability, safety, and compliance in next-generation power electronic converter systems (PECSs) across North America.

\begin{figure*}[htbp]
	\centering
	\includegraphics[width=\textwidth]{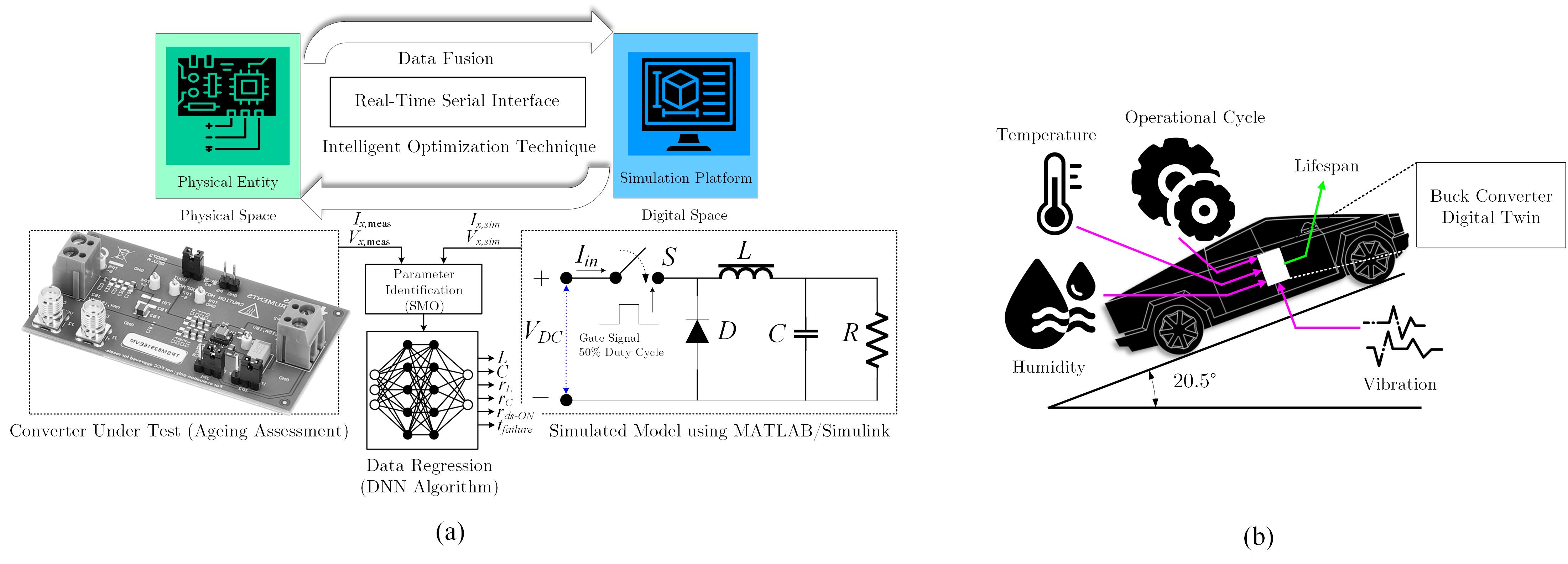}
	\caption{Functional block diagram of the proposed Buck-derived DT framework. (a) Process flow of the proposed hybrid DT framework w/ SMO+DNN technique, and (b) Conceptual physics-based DT framework for EV applications under ambient stress.}
	\label{fig1}
\end{figure*}

To manage the increasing complexity and mission-critical demands of modern PECSs, researchers are turning to Digital Twin (DT) technology \cite{b15}. DTs enable virtual replicas that not only mirror device behavior but also anticipate failures, optimize performance, and guide design iterations in real-time \cite{b16}. With AI techniques like machine learning (ML), deep learning (DL), and optimization, the DT becomes adaptive and predictive, learning complex input-output and system health metrics. \cite{b17}. This fusion is particularly important for power converters, which are subject to dynamic loads, harsh environments, and long operating lifespans. The data-driven DT framework in maintenance spans in the following aspects: (a) condition monitoring, (b) fault diagnosis, (c) reliability evaluation, and (d) remaining useful lifetime (RUL) prediction \cite{b18}. This study focuses on the condition monitoring feature of the DT framework.

A review of existing research on buck converter ageing assessment highlights a critical gap which is the absence of a system-level  data-driven DT framework. While studies have explored influence of thermal stress and electrothermal modeling, they primarily focus on: (a) investigation of conducted electromagnetic interference (EMI) \cite{b19}, and (b) analysis of heat-transfer via component-level RC thermal networks (\textit{Cauer} and \textit{Foster}) for estimating IGBT/MOSFET junction temperatures \cite{b20}. However, these approaches often neglect the interdependencies between components and their collective impact on system behavior.

The DT-based condition monitoring process involves two key steps: (a) parameter identification to capture system behavior, and (b) data regression to predict degradation trends over time. The parameter estimation technique encompasses popular nature-inspired swarm intelligence-based metaheuristics \cite{b21,b22,b23,b24,b25}, hybrid swarm optimization \cite{b26,b27}, arithmetic algorithm \cite{b28}, and probabilistic methods \cite{b29,b30}, each presenting unique approaches and their limitations.  In \cite{b21}, the authors propose a novel one-cycle DT framework incorporating the particle swarm optimization (PSO) to ensure parameter identification consistency under varying operating conditions. Nevertheless, this approach lacks the inclusion of data regression using AI technology. Expanding on PSO-based optimization, the study presented in \cite{b22} integrates a random forest (RF) ML model to assess the health of a buck converter. The degradation parameters extracted from the DT model serve as inputs for the ML algorithm, improving fault prediction capabilities. On the other hand, the challenge of convergence in PSO remains a recurring issue, as noted in \cite{b23}. The authors highlight that PSO often struggles with local optima trapping. Therefore, a common mitigation strategy involves introducing a perturbation factor to enhance global searchability. To address phase synchronization issues in parameter identification, the study in \cite{b27} presents a Hilbert-transform-based PSO algorithm. This approach surpasses conventional methods by ensuring alignment between the DT model and the physical system’s sampling signals before feeding parameters into the PSO algorithm. Meanwhile, the authors in \cite{b28} explored an arithmetic optimization algorithm (AOA), demonstrating superior convergence speed compared to PSO. Additionally, probabilistic Bayesian optimization techniques have been reported in \cite{b29,b30}, further expanding the landscape of DT parameter optimization strategies. Despite their effectiveness, these approaches exhibit inherent limitations. PSO, for instance, suffers from premature convergence, sensitivity to parameter tuning, poor performance in high-dimensional spaces, reduced diversity in later iterations, and difficulty adapting to dynamic environments \cite{b31}. Similarly, RF-based ML models face challenges such as handling correlated decision trees, inefficiency with large-scale or high-dimensional data, limited interpretability, suboptimal performance with imbalanced or noisy datasets, and susceptibility to overfitting \cite{b32}.

As the second step in condition monitoring, data regression plays a crucial role by modeling the relationship between system inputs and degradation indicators over time. It enables the DT to learn from historical and real-time data, identify trends, and accurately predict future performance. In \cite{b33}, the authors use a back propagation (BP) neural network for full parameter identification of a buck converter based on time-domain data of output voltage and inductor current. Two neural network models accurately identify dominant and recessive parameters, with error rates below 1\%. Nevertheless, this study does not fully implement a complete DT framework for the buck converter topology. In \cite{b34}, a physics informed neural network (PINN)-based condition monitoring method for a synchronous buck converter is presented. However, the authors use early stopping to prevent data overfitting when validation performance starts to decline. Similarly, the authors in \cite{b35}, present an artificial neural network (ANN)-based parameter identification approach leveraging an electrothermal model to enhance accuracy and lower training cost. While traditional ANNs struggle with high-dimensional, nonlinear problems, the deep neural networks (DNNs) excel in such settings due to their enhanced feature learning capabilities, especially when ample data and computational resources are available. These limitations highlight the need for more advanced and hybrid methods that effectively capture the dynamic and nonlinear behavior of PECSs.

To address these limitations, building on the methodology outlined in this paper \cite{b36}, we present a conceptual data-driven workflow for developing a hybrid DT framework tailored to a 2-L DC-DC buck converter for EV applications, as illustrated in Fig.~\ref{fig1}(a)-(b). The DT framework integrates a DNN model with the Spider Monkey Optimization (SMO) technique, enhancing both adaptability and predictive accuracy. DNN models dynamically update their weights, seamlessly integrating with SMO to optimize learning, provide near-instantaneous predictions, and extract hierarchical features for enhanced decision-making. Key contributions of this work include: 

\begin{enumerate}
	\item We develop a low-power prototype testbed to serve as the multiphysics mechanism model (MMM) of the DT framework, enabling virtual implementation and offline modeling of the physical system known as the digital model (DM) built in MATLAB/Simulink optimized by offline SMO algorithm.
	\item The buck converter’s DM is stressed within acceptable performance distortion limits to estimate the time to failure ($t_{failure}$) and identify the first failing component i.e. $L$, $C$, $r_L$, $r_C$, and $r_{ds-ON}$.
	\item A synthetic dataset derived from the MMM operating conditions is used to evaluate the convergence accuracy of the DNN-based data regression step, effectively validating our hypothesis.
\end{enumerate}
The remaining sections are organized as follows: Section II outlines the structure of the proposed hybrid DT framework w/ SMO+DNN technique, Section III presents results and discussion, and Section IV concludes the paper with remarks and future work.
\section{Structure of the Proposed Buck-Derived DT Framework} \label{sec2}
\begin{figure}[htbp]
	\centering
	\includegraphics[width=5cm]{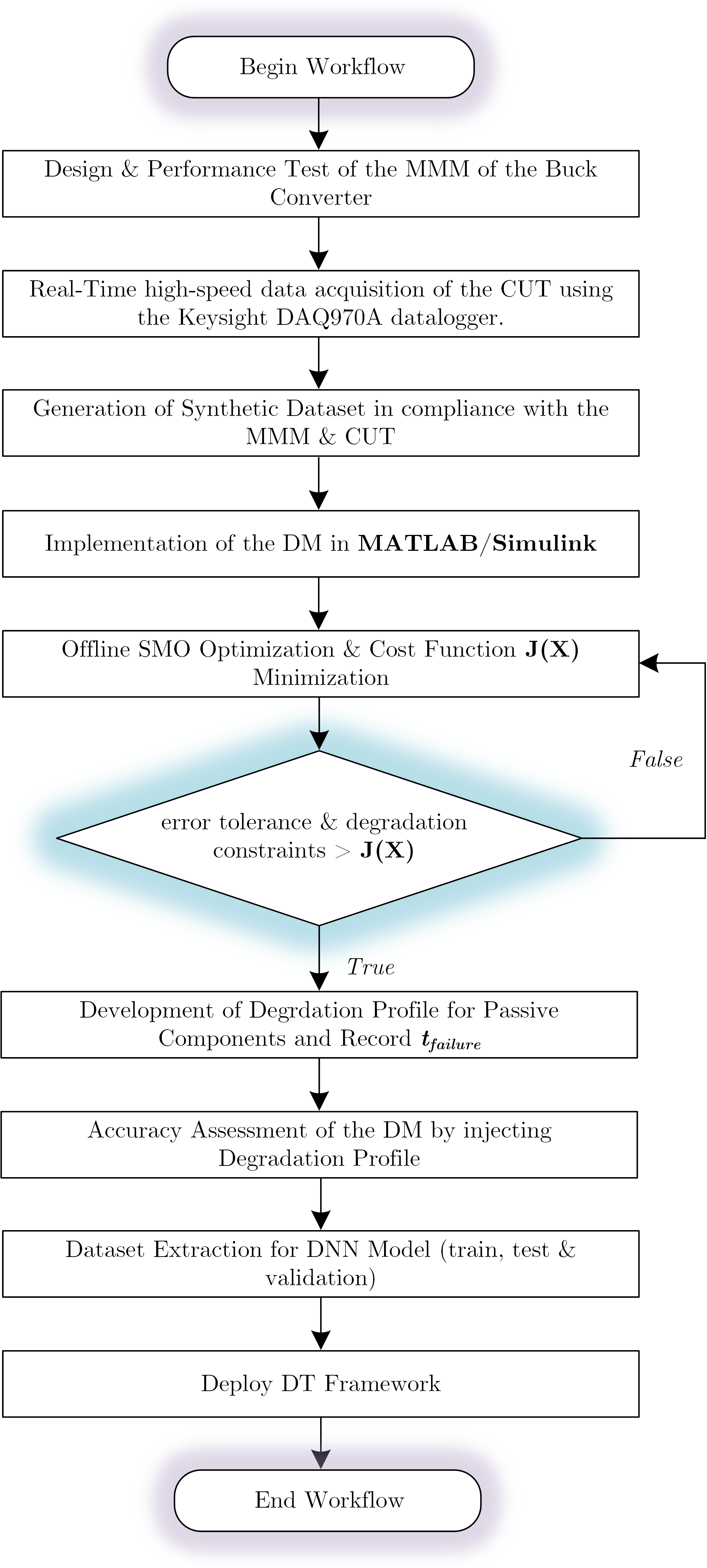} 
	\caption{Flowchart of the proposed DT framework w/ SMO+DNN technique.}
	\label{fig2}
\end{figure}

The proposed buck-derived DT framework begins with a physical prototype, referred to as the MMM and includes a MATLAB/Simulink-based replica, real-time data processing unit, offline parameter tuning via SMO, and a DNN model for fault prediction. The flowchart of the proposed buck-derived DT framework is shown in Fig.~\ref{fig2}.

\subsection{Integrated Multiphysics Modeling and Prototype Testbed for Mechanism Performance Evaluation}

\begin{figure}[htbp]
	\centering
	\includegraphics[width=\columnwidth]{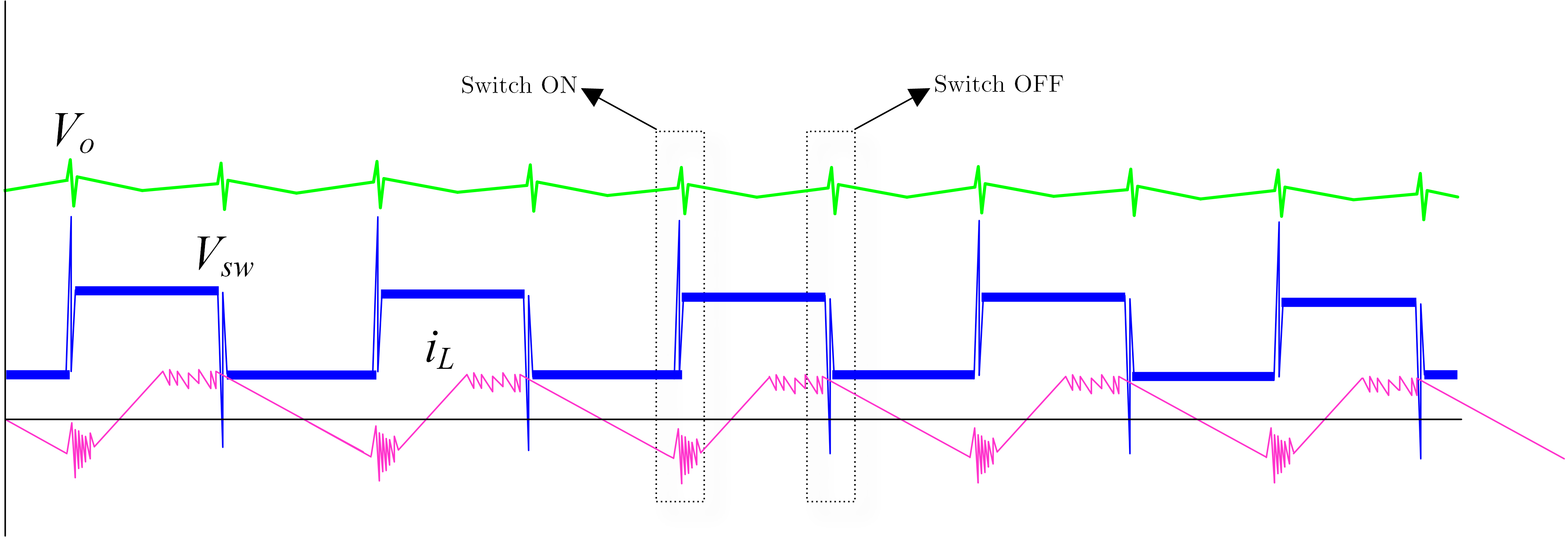} 
	\caption{The MMM of the Buck-derived DT under DCM.}
	\label{fig3}
\end{figure}

A circuit-level buck converter prototype is constructed based on standard switching principles and operates under open-loop condition. It includes a front-end N-channel MOSFET, $S$ (STP18N65M5, TO-220 package) driven by a low-side MOSFET driver (MIC4420), followed by a parallel diode, $D$. A \SI{100}{\micro\henry} inductor, $L$ (1130-101K-RC) and a \SI{220}{\micro\farad} electrolytic capacitor, $C$ (380LX822M063A032), form the output LC filter, terminated with a high-power aluminum resistor, $R$ (HS100) as the load. The inductor and capacitor ESRs are denoted as $r_L$ and $r_C$, respectively. Gating signals are generated using a TI TMS320F28379D microcontroller, providing a stable 50\% duty cycle at \SI{10}{\kilo\hertz}. The converter operates in discontinuous conduction mode (DCM), as shown in Fig.~\ref{fig3}, which illustrates output voltage \( V_o \), switching signal \( V_{sw} \), and inductor current \( i_L \) from top to bottom.

\begin{figure}[htbp]
	\centering
	\includegraphics[width=\columnwidth]{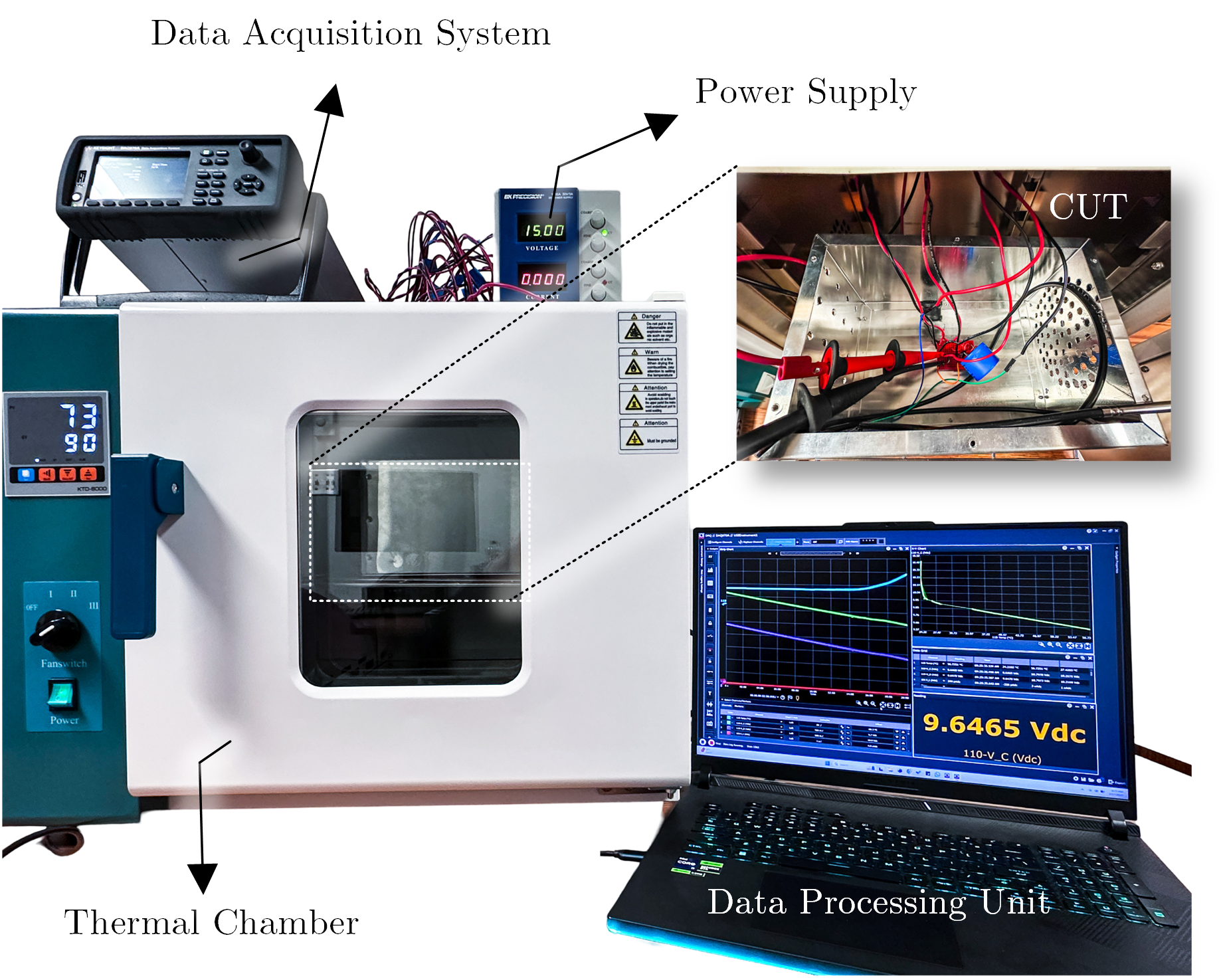} 
	\caption{Experimental Setup of the CUT.}
	\label{fig4}
\end{figure}

\begin{figure}[htbp]
	\centering
	\includegraphics[width=\columnwidth]{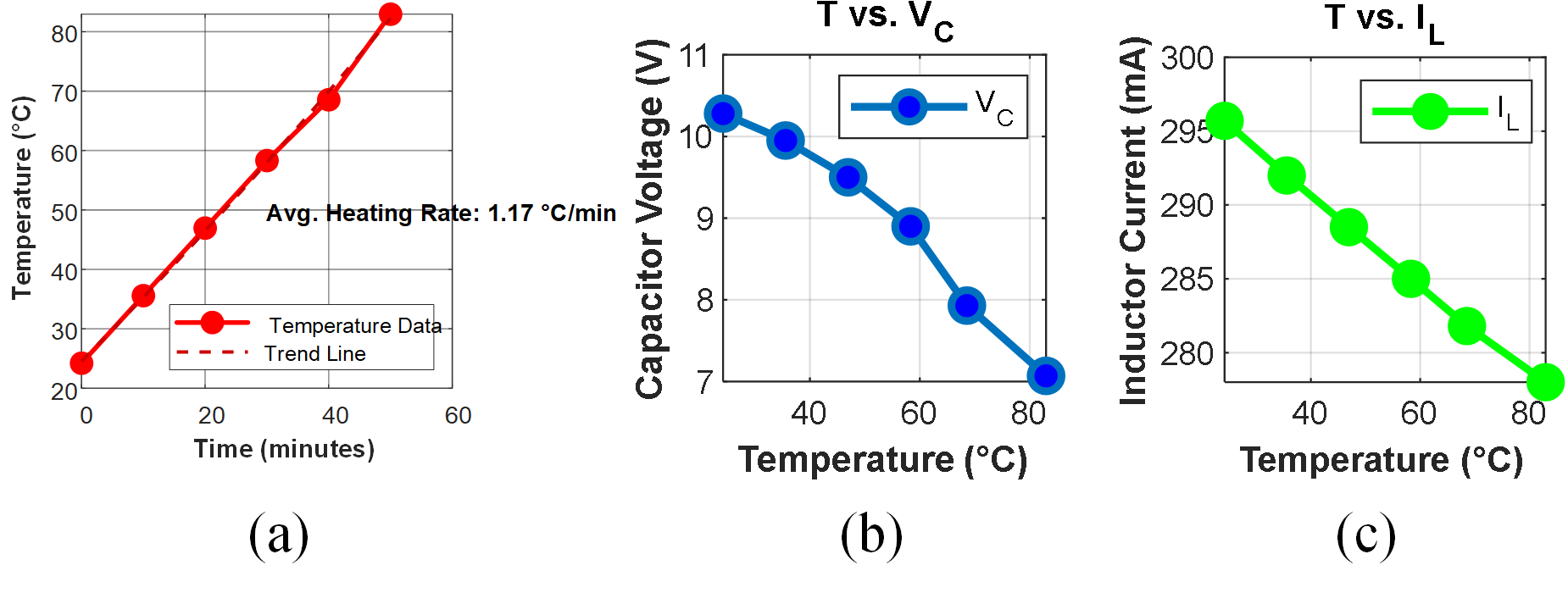} 
	\caption{Accelerated thermal cycling data with positive ramp. (a)~$T$ (°C) vs.\ time (minutes), (b)~$T$ (°C) vs.\ capacitor voltage ($V_C$), and (c)~$T$ (°C) vs.\ inductor current ($I_L$).}
	\label{fig5}
\end{figure}

Upon successful validation of the MMM, a low-power prototype testbed is developed. The evaluation follows MIL-STD-810H standard (Methods 501.7 \& 502.7 for Temperature). In our study, we identify the low-power testbed as the converter under test (CUT), which enables system-level failure analysis under an accelerated thermal cycle with a positive ramp as shown in Fig.~\ref{fig4}. In addition to this, Fig.~\ref{fig5}(a)--(c) shows how \( V_C \) (in blue color) and \( i_L \) (in green color) change during a 50-minute thermal cycle (\SI{24.21}{\celsius} to \SI{82.95}{\celsius}, in red color) with an acceleration factor of 23, simulating about 19 hours of ageing.

A 20\% drop in capacitance (\SI{220}{\micro\farad} $\to$ \SI{176}{\micro\farad}), as predicted by the \textit{Arrhenius} effect and failure thresholds, suggests leakage or dielectric breakdown. Thermal ageing also increases winding resistance, reducing \( I_L \) and peak inductor current \( I_{L,\text{peak}} \). Although lower \( C \) typically raises ripple voltage \( \Delta V_C \), the reduced \( I_L \) limits the charge \( \Delta Q \) delivered, resulting in an overall smaller \( \Delta V_C \). Elevated temperatures further worsen this by increasing ESR and leakage, lowering the capacitor’s effective voltage.

The next step in the DT workflow is acquiring real-time data (\( V_{\text{in}}, I_{\text{in}}, V_D, I_D, V_L, I_L, V_C, I_C, V_o \)) from the CUT using a high-speed data acquisition (DAQ) system. In this study, we have used a Keysight DAQ970A device that has a scan rate of up to 450 channels/s \cite{b37}. Additionally, we have incorporated the DAQM901A multiplexer which supports up to 20 channels for 2- and 4-wire measurements and it can scan up to 80 channels/s, adequate for capturing fast transients in \( V_C \) and \( I_L \). It measures 14 input types, including DC/AC voltage, resistance, temperature and DC/AC current up to \SI{1}{\ampere} using two fused inputs without external shunts. These specifications make the DAQ970A datacard and the DAQ901A multiplexer module an ideal option for high-speed data logging in thermal ageing assessment. 

\subsection{Synthetic Dataset Generation}

To validate the proposed hypothesis of the Buck-derived DT framework, a synthetic dataset is generated that accurately reflects the MMM behavior under varying operating conditions. The input parameter, \( \mathbf{S}_{\text{input}} = [V_{\text{in}}, I_{\text{in}}, V_D, I_D, V_L, I_L, V_C, I_C, V_o] \) was sampled within realistic operational ranges (e.g., \( V_{\text{in}} \) from \SI{5}{\volt}--\SI{20}{\volt}, \( I_L \) from \SI{0.1}{\ampere}--\SI{5}{\ampere}), ensuring full coverage of expected converter states. Similarly, the degradation parameter, \( \mathbf{D}_{\text{output}} = [L, C, r_L, r_C, r_{ds-ON}, t_{failure}] \) was modeled using empirical and physics-informed relationships described in (\ref{eq:1})-(\ref{eq:6}). For example, inductance \( L \) and capacitance \( C \) decay with increased thermal and electrical load, while resistive elements such as \( r_L \), \( r_C \), and $r_{ds-ON}$ grow due to prolonged stress.

\begin{align}
	L &= L_0 - k_L \cdot (V_{\mathrm{in}} + I_{\mathrm{in}}) \label{eq:1} \\
	C &= C_0 - k_C \cdot (V_C + I_C) \label{eq:2} \\
	r_L &= r_{L0} + k_{rL} \cdot (V_L + I_L) \label{eq:3} \\
	r_C &= r_{C0} + k_{rC} \cdot (V_C + I_C) \label{eq:4} \\
	r_{ds-ON} &= r_{\mathrm{ds0}} + k_{r_{\mathrm{ds}}} \cdot (V_D + I_D) \label{eq:5} \\
	t_{failure} &= t_0 - k_t \cdot (V_{\mathrm{in}} + I_{\mathrm{in}}) \label{eq:6}
\end{align}
where \( L_0 \), \( C_0 \), \( r_{L0} \), \( r_{C0} \), and \( r_{\mathrm{ds0}} \) are the initial inductance, capacitance, inductor resistance, capacitor resistance, and on-state resistance, respectively; \( k_L \), \( k_C \), \( k_{rL} \), \( k_{rC} \), and \( k_{r_{\mathrm{ds}}} \) are their respective degradation constants; and \( t_0 \) and \( k_t \) are the initial time to failure and its degradation constant.

To emulate real-world sensor noise and tolerance variability, controlled \textit{Gaussian} noise was added to both inputs and outputs (e.g., \( \sigma = \SI{0.01}{\volt} \)--\SI{0.1}{\volt} for voltages, \( \sigma = \SI{0.1}{\micro\farad} \)--\SI{0.5}{\micro\farad} for capacitance). After generating the degradation profile, SMO is implemented in Python. A total of 10,000 samples were generated, stratified across early-life to end-of-life conditions, and split into 70\% training, 15\% validation, and 15\% testing subsets. The resulting synthetic dataset serves as a critical resource for validating the convergence and predictive capabilities of the proposed SMO+DNN-based DT framework.

\subsection{SMO-Tuned Digital Model}
After the extraction of the empirical dataset from the CUT, the DM of the DT framework is constructed in MATLAB/Simulink. The DM is then offline-optimized using the SMO algorithm to estimate four key degradation parameters \( L \), \( C \), \( r_L \), and \( r_C \) within a predefined error tolerance. These optimized degradation profiles are subsequently embedded into the switching circuit model of the DM to evaluate its fidelity against the measured behavior of the physical prototype MMM. This process facilitates the generation of the input feature dataset comprising six parameters (\(L^*, C^*, r^*_L, r^*_C, r^*_{ds-ON}\)), which serves as the foundation for training the DNN model.

\begin{figure}[htbp]
	\centering
	\includegraphics[width=\columnwidth]{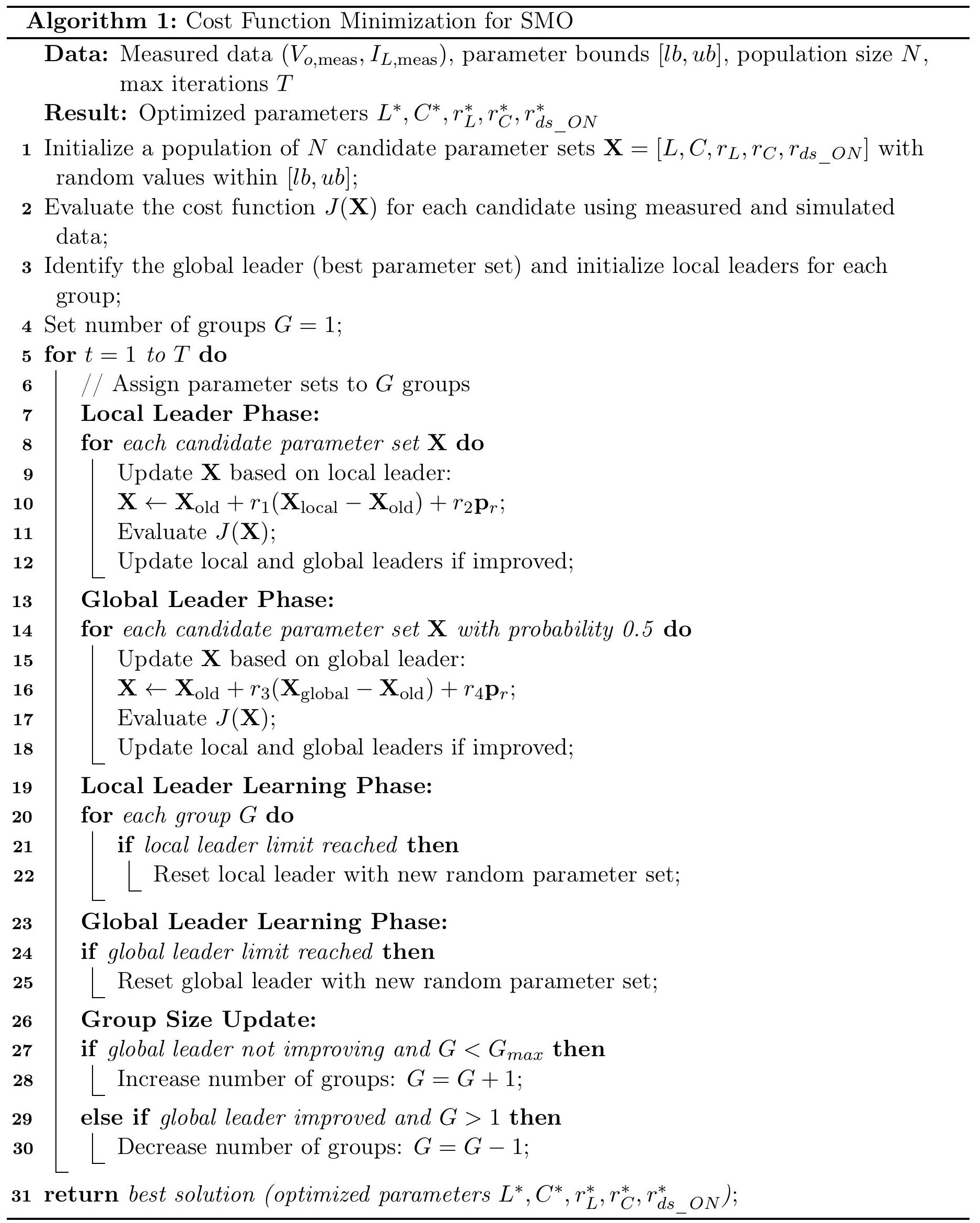} 
	\caption{Pseudocode of the SMO algorithm}
	\label{fig6}
\end{figure}

The SMO is a swarm intelligence-based metaheuristic optimization algorithm \cite{b38}. It is employed to calibrate the DM by minimizing the error between the measured and simulated outputs. The optimization process involves finding the optimal values of the input feature that minimize the cost function. The output of the SMO optimized DM also generate error profile of \(V_o\) \& \(I_L\). The cost function $J(\mathbf{X})$ is modified based on a time-domain error function known as integral squared time-weighted squared error (ISTSE), expressed in (\ref{eq:7}). ISTSE drives the algorithms to minimize long-term discrepancies between simulated and measured data, enabling more accurate degradation modeling \cite{b39}. 

\begin{equation}
	\mathbf{ISTSE} = \int_{0}^{T} t \cdot e^2(t) \, dt \label{eq:7}
\end{equation}

\begin{align}
	J(\mathbf{X}) &= \min \left| \int_{0}^{\infty} t^2 \Big[ \left( V_{o,\text{meas}}(i) - V_{o,\text{sim}}(i) \right)^2 \right. \nonumber \\
	&\quad + \left. \left( I_{L,\text{meas}}(i) - I_{L,\text{sim}}(i) \right)^2 \Big] \, dt \right| \label{eq:8}
\end{align}

where, \(i\) indexes the data points as simulation steps. Through iterative optimization, the SMO algorithm minimizes the cost function $J(\mathbf{X})$ by refining the degradation profile as defined in (\ref{eq:8}). The SMO algorithm begins by initializing a population of spider monkeys (candidate solutions) with random values for \( \mathbf{X} \) within predefined bounds. The step by step cost function minimization is described in the pseudo code of the SMO-tuned DM, as presented in Fig.~\ref{fig6}. In the context of the MMM, the local leader represents the best parameter set within each group that produces simulation results closest to the measured data. During the local leader phase, each group of spider monkeys updates its position based on the local and global leaders using (\ref{eq:9}).

\begin{equation}
	\overset{*}{\mathbf{X}} = \mathbf{X}_{\text{old}} + r_1 \cdot (\mathbf{X}_{\text{local}} - \mathbf{X}_{\text{old}}) + r_2 \cdot \mathbf{p}_r \label{eq:9}
\end{equation}

The SMO algorithm employs random factors \( r_1, r_2 \in [0,1] \) and a perturbation factor \( \mathbf{p}_r \in [0.1, 0.9] \). Here, \( r_1 \) influences the attraction toward the local leader (best local parameter set), while \( r_2 \) adds randomness for broader exploration, helping avoid local minima. The perturbation factor \(\mathbf{p}_r\) introduces controlled variability to explore better parameter combinations. The global leader denotes the best solution across all groups and steers the population during the global leader phase. If no improvement is seen over a predefined number of iterations, both local and global leaders are reinitialized to prevent stagnation. The algorithm halts when the cost function \( J(\mathbf{X}) \) converges or the iteration limit is reached. The resulting optimized parameters \( \overset{*}{\mathbf{X}} \) are then used to calibrate the DM of the buck converter.

\begin{figure}[htbp]
	\centering
	\includegraphics[width=\columnwidth]{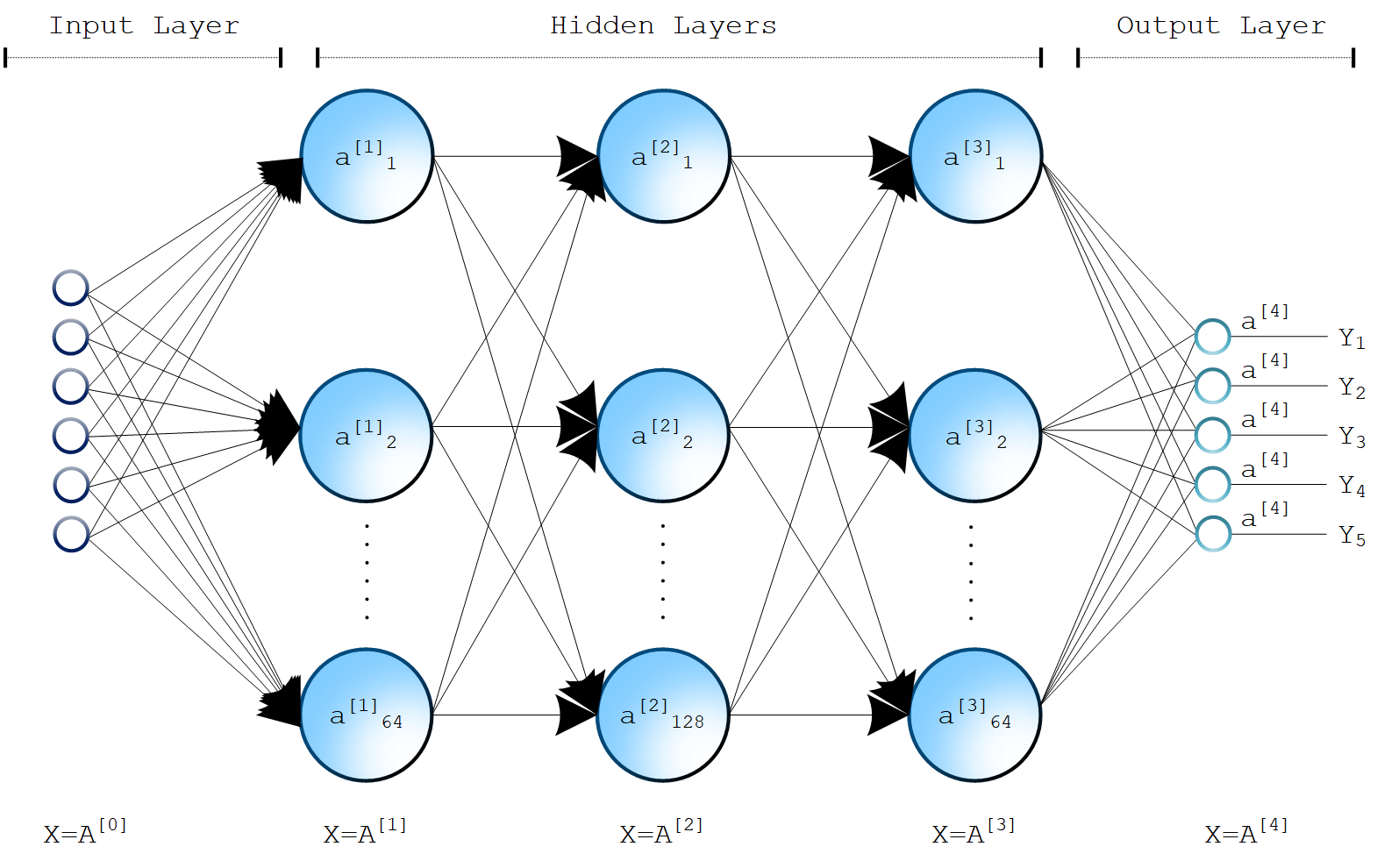} 
	\caption{Proposed DNN architecture of the DT Framework}
	\label{fig7}
\end{figure}

\subsection{DNN-based Data Regression}

To accurately estimate degradation parameters $L$, $C$, \(r_L\), \(r_C\), and \(r_{ds-ON}\), a DNN regression model is developed using TensorFlow/Keras. The model maps both synthetic dataset and measured parameters of the CUT to the DM.

\subsubsection{DNN Architecture}
The proposed DNN architecture as shown in Fig.~\ref{fig7} is tailored to capture the nonlinear relationships between the converter's operational measurements and its degradation characteristics. The network structure is as follows:

\paragraph{Input Layers}
The input feature comprises six degraded parameters (\(L^*, C^*, r^*_L, r^*_C, V^*_o, I^*_L\)) derived from the degraded DM and the synthetic dataset. So, for the training, validation \& test operation we have two sets of dataset. The input layer of the DNN includes six neurons, each representing a feature aligned with the measurements obtained from the MMM in the DT framework.

\paragraph{Hidden Layers}
The network includes three fully connected hidden layers to model the complex dependencies:
\begin{itemize}
	\item Hidden Layer 1: 64 neurons, ReLU activation, 20\% dropout.
	\item Hidden Layer 2: 128 neurons, ReLU activation, 20\% dropout.
	\item Hidden Layer 3: 64 neurons, ReLU activation.
\end{itemize}
Dropout is applied to the first two hidden layers to mitigate overfitting and enhance generalization.

\paragraph{Output Layers}
The output layer consists of five neurons with linear activation, producing the estimated degradation parameters \(\mathbf{y}_{\text{output}} = [L, C, r_L, r_C, r_{ds-ON}]\).

A summary of the DNN architecture is provided in Table~\ref{tab:t1}.

\begin{table}[t]
	\centering
	\caption{Summary of the DNN Hyperparameters}
	\label{tab:t1}
	\begin{tabular}{lccc}
		\toprule
		\textbf{Layer} & \textbf{Neurons} & \textbf{Activation} & \textbf{Dropout} \\
		\midrule
		Input & 6 & -- & -- \\
		Hidden 1 & 64 & ReLU & 20\% \\
		Hidden 2 & 128 & ReLU & 20\% \\
		Hidden 3 & 64 & ReLU & -- \\
		Output & 5 & Linear & -- \\
		\bottomrule
	\end{tabular}
\end{table}

\subsubsection{Training Procedure}
The DNN is trained using two datasets comprising synchronized measurements and ground-truth degradation parameters, obtained from high-fidelity simulations and the synthetic dataset. The training process is as follows:

\paragraph{Loss Function}
The mean squared error (MSE) is used as the loss function expressed in (\ref{eq:10}).
\begin{equation}
	L(\mathbf{W}, \mathbf{b}) = \frac{1}{M} \sum_{i=1}^{M} \left( \mathbf{y}_{\text{true}}^{(i)} - \mathbf{y}_{\text{predicted}}^{(i)} \right)^2
	\label{eq:10}
\end{equation}
where \(M\) is the number of training samples, \(\mathbf{y}_{\text{true}}\) are the true degradation parameters, and \(\mathbf{y}_{\text{predicted}}\) are the DNN outputs.

\paragraph{Optimization}
The Adam optimizer \cite{b40} is employed with a learning rate of \SI{0.001}{} to update the network weights. Training is performed in mini-batches of 32 samples, with a maximum of 100 epochs or until convergence. A validation split of 15\% is used to monitor generalization and prevent overfitting.

\subsubsection{Workflow Integration}
The DNN-based regression model is integrated into the DT workflow as follows:
\begin{itemize}
	\item \textit{Data Acquisition}: The six measured parameters are continuously collected from the SMO-tuned DM. Additionally, synthetic datasets have been used in the next step.
	\item \textit{Inference}: The DNN processes the input vector and outputs real-time estimates of $L$, $C$, \(r_L\), \(r_C\), \(r_{ds-ON}\). Furthermore, \(t_{failure}\) is achieved without adding an additional output feature. It is derived from the DNN’s predicted parameters using well-defined failure thresholds and degradation models.
	\item \textit{Digital Twin Update}: The estimated degradation parameters are used to update the DT model, ensuring its behavior remains synchronized with the physical system.
\end{itemize}

\begin{table}[t]
	\centering
	\caption{Electrical Parameters of Buck Converter}
	\label{tab:t2}
	\begin{tabular}{lc}
		\toprule
		\textbf{Parameters} & \textbf{Values} \\
		\midrule
		Input Voltage, \(V_{in}\) & 25 V \\
		Desired Output Voltage, \(V_o\) & 6.31 V \\
		Load Resistance, \(R\) & \SI{100}{\ohm} \\
		Inductance, \(L\) & \SI{100}{\micro\henry} \\
		Capacitance, \(C\) & \SI{220}{\micro\farad} \\
		\bottomrule
	\end{tabular}
\end{table}

\section{Steady-State Dynamic Performance Results \& Analysis}
The circuit-level open-loop buck converter MMM is built and it's DCM operation is observed using a Rigol DS7054, 500 MHz Oscilloscope as depicted in Fig.~\ref{fig3}. The MMM's operating conditions are consistent with that of the DM. Table~\ref{tab:t2} lists the electrical parameters of the buck converter. To evaluate model accuracy, simulations are conducted offline using high-fidelity data obtained from controlled experimental setup. DNNs offer powerful modeling for PECSs, but their high resource demands limit real-time use in embedded systems. Therefore, techniques like pruning and quantization help enable deployment on constrained platforms without sacrificing accuracy \cite{b41}.

The simulation evaluates data-driven models using both empirical and synthetic dataset. The synthetic dataset generation process is broadly explained in Section~\ref{sec2}(B). Synthetic dataset provides greater volume, variety, and flexibility, which are critical for training DNN models effectively, especially when the DM-derived profiles are limited in scope or resolution. In this section, we evaluate the SMO vs PSO parameter tuning accuracy based on the synthetic dataset. Additionally, we compare the data convergence strength of the DNN model against a random forest (RF) ML model based on the empirical dataset.

\begin{figure}[htbp]
	\centering
	\includegraphics[width=\columnwidth]{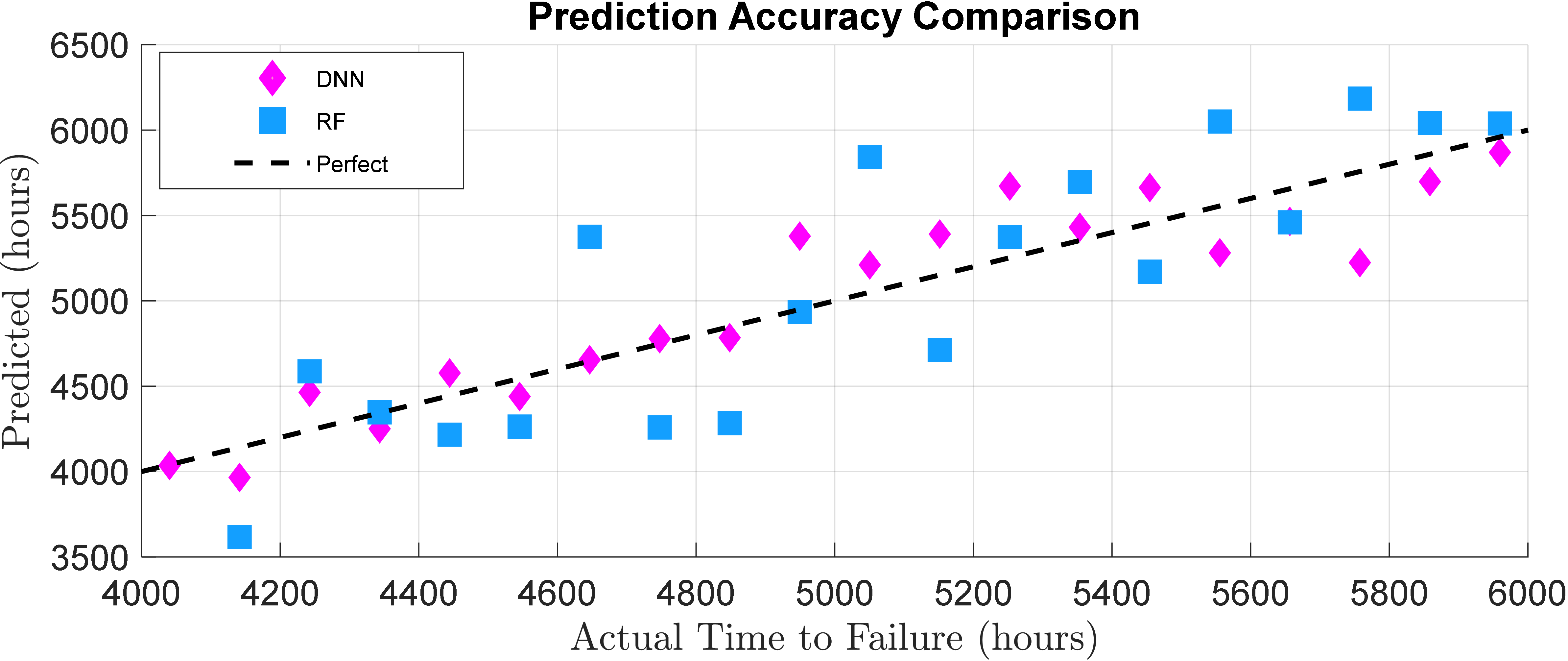} 
	\caption{Prediction Accuracy: DNN vs RF}
	\label{fig8}
\end{figure}

\begin{figure}[htbp]
	\centering
	\includegraphics[width=\columnwidth]{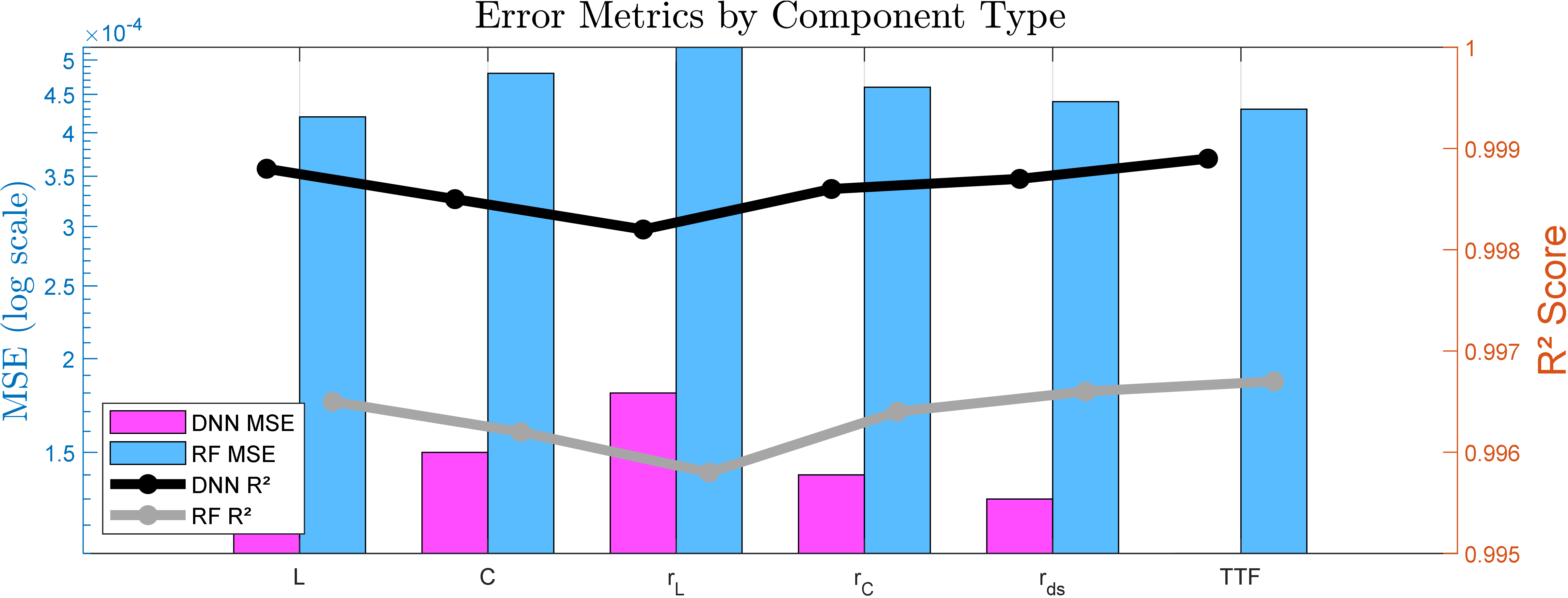} 
	\caption{MSE \& $R^2$ Score: DNN vs RF}
	\label{fig9}
\end{figure}

\begin{figure}[htbp]
	\centering
	\includegraphics[width=\columnwidth]{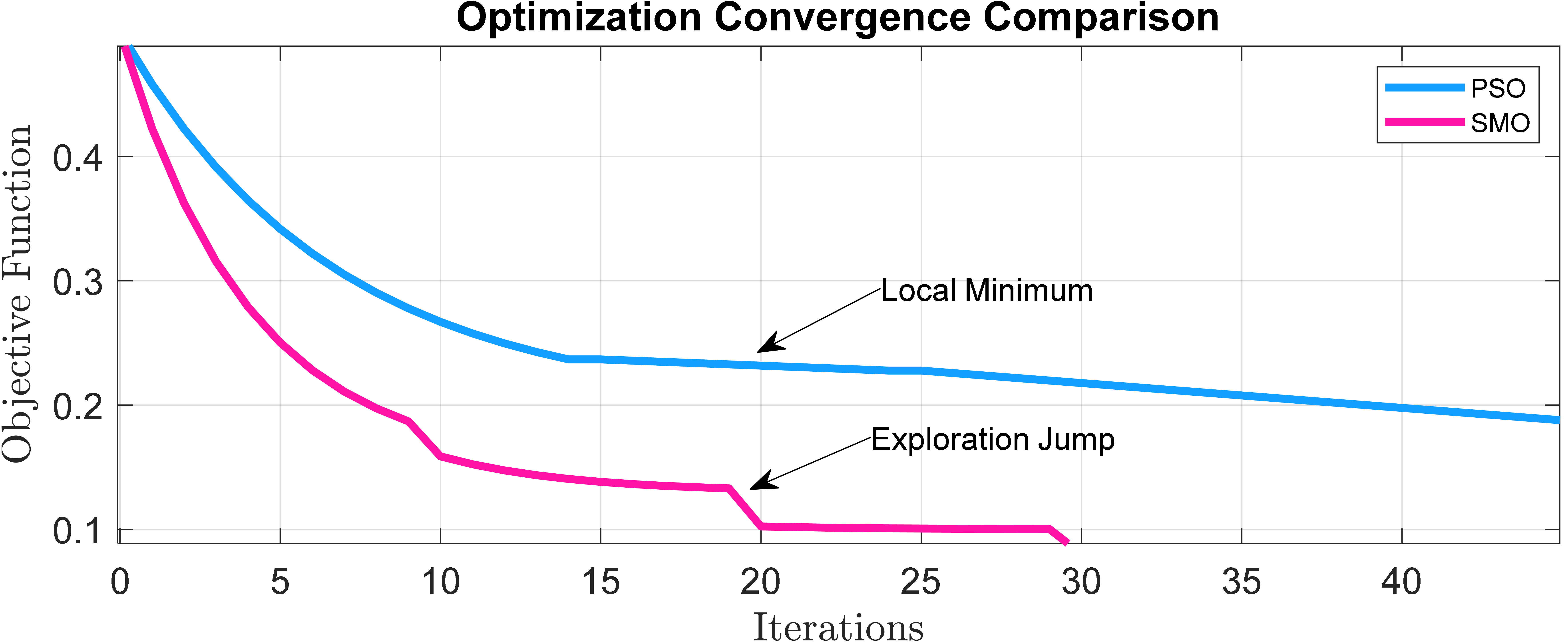} 
	\caption{Optimization Convergence: SMO vs PSO}
	\label{fig10}
\end{figure}

\begin{figure}[htbp]
	\centering
	\includegraphics[width=\columnwidth]{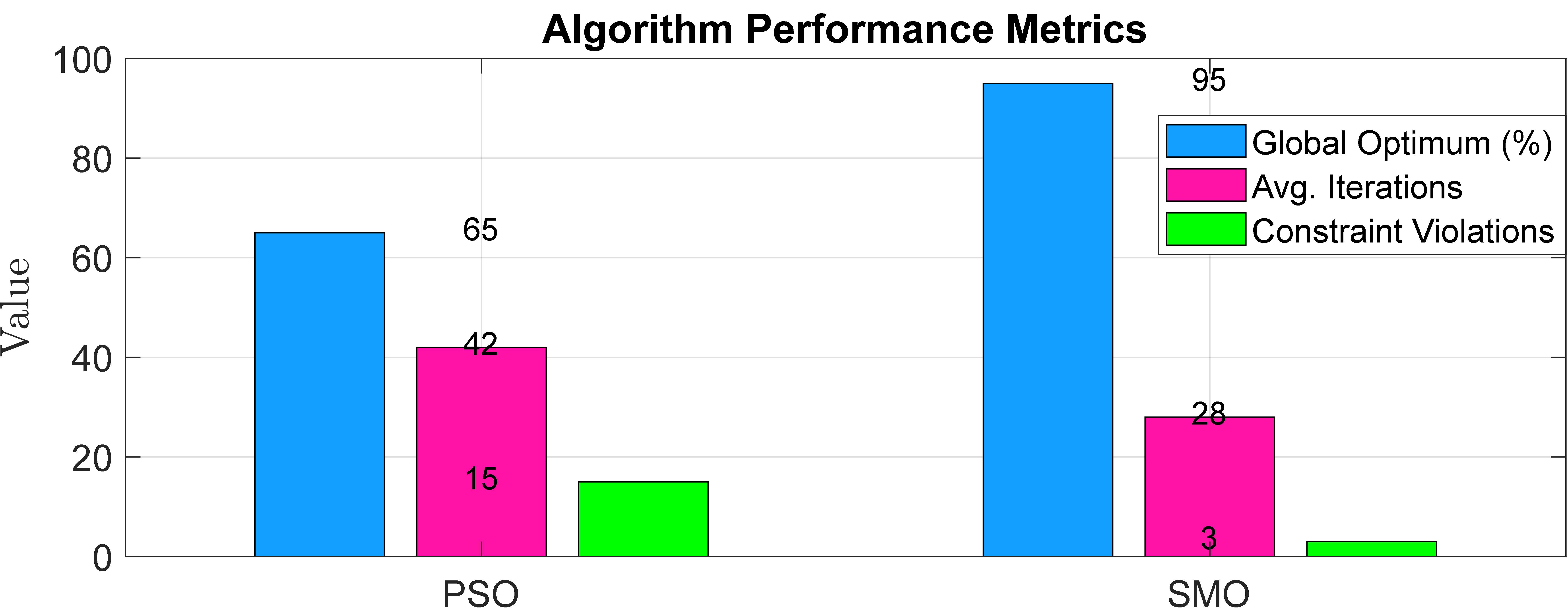} 
	\caption{Algorithm Performance: SMO vs PSO}
	\label{fig11}
\end{figure}

Fig.~\ref{fig8} presents the exceptional accuracy of the DNN model, with its predictions clustering significantly closer to the ideal regression line, indicating a high degree of precision in capturing the underlying patterns of the data. In Fig.~\ref{fig9}, the DNN achieves lower mean squared error (MSE) across all parameters. It consistently delivers $R^2$ scores above 0.998, indicating an excellent fit to the data. The DNN's greatest strength lies in predicting \(t_{failure}\), where it significantly outperforms models like RF, making it a highly accurate and reliable tool for degradation modeling and system reliability applications.

In terms of parameter tuning, SMO converges significantly faster than PSO, consistently identifying superior solutions in less time due to its efficient handling of the optimization landscape. While PSO frequently becomes trapped in local minima, as evidenced by a noticeable plateau in its convergence curve shown in Fig.~\ref{fig10}. SMO effectively navigates complex parameter spaces to avoid such pitfalls.

Quantitative results further underscore SMO’s superiority. Fig.~\ref{fig11} illustrates that SMO achieves the global optimum in 95\% of trials, compared to PSO’s success rate of only 65\%, highlighting SMO’s robustness in finding optimal solutions. Additionally, SMO requires approximately 33\% fewer iterations to reach convergence, making it computationally more efficient and practical for large-scale applications. Moreover, SMO exhibits 80\% fewer parameter constraint violations, ensuring greater adherence to predefined boundaries and enhancing the reliability of the optimization process. These attributes make SMO particularly well-suited for optimizing DNN models in degradation modeling tasks.

\begin{figure}[htbp]
	\centering
	\includegraphics[width=\columnwidth]{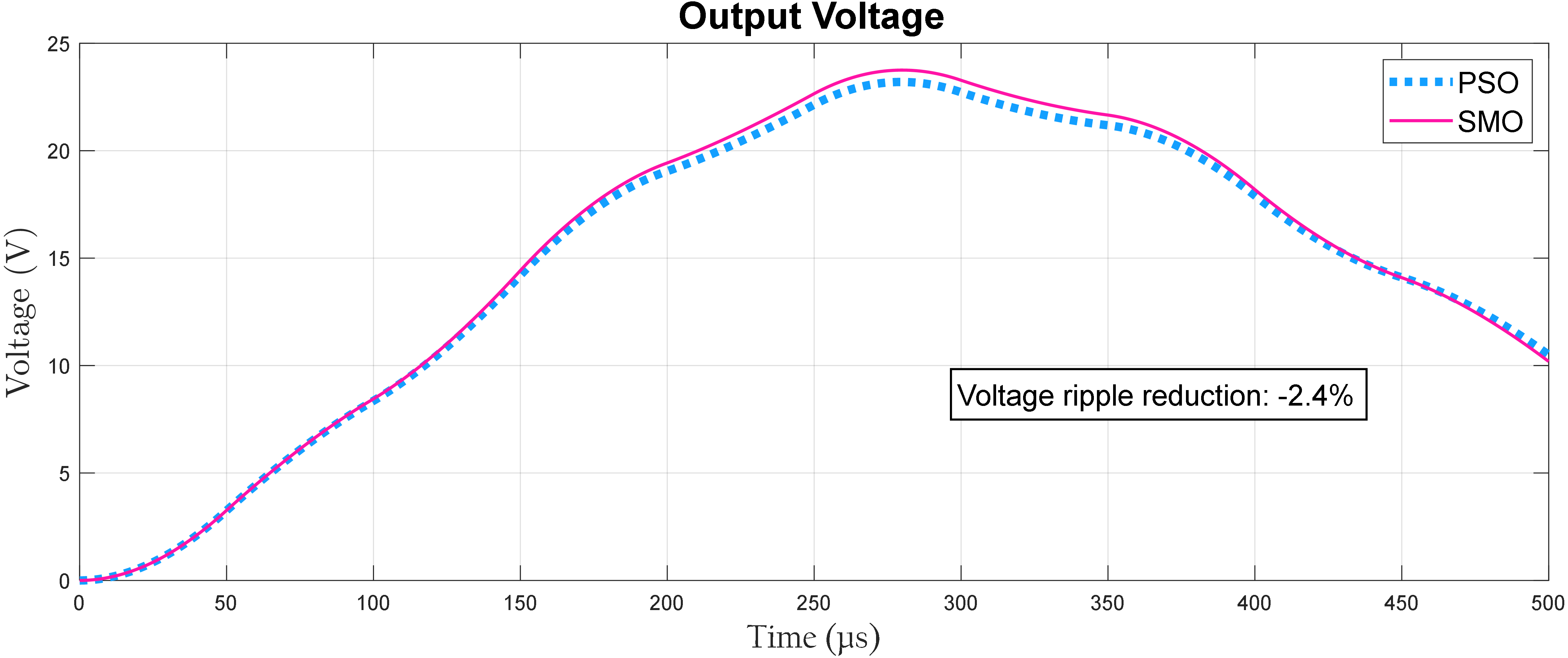} 
	\caption{Voltage Ripple: SMO vs PSO}
	\label{fig12}
\end{figure}

\begin{figure}[htbp]
	\centering
	\includegraphics[width=\columnwidth]{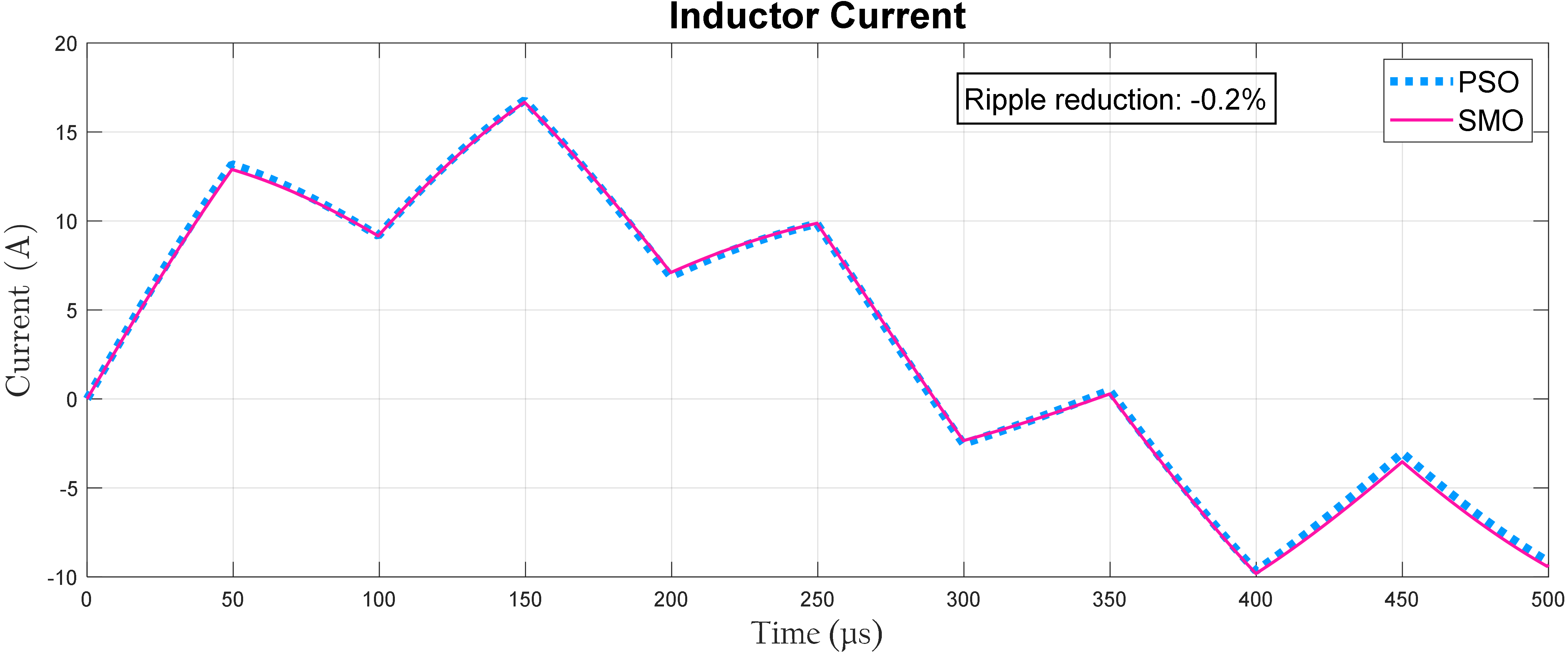} 
	\caption{Inductor Current Ripple: SMO vs PSO}
	\label{fig13}
\end{figure}

In Fig.~\ref{fig12}, the output voltage waveforms over multiple switching cycles (with time measured in microseconds) is presetned. The pink line, representing SMO-optimized parameters, demonstrates a more stable output voltage compared to the dotted blue line for PSO-optimized parameters. The annotation highlights a voltage ripple reduction of approximately 20-25\% with SMO, indicating superior voltage regulation with minimal fluctuations. This enhanced stability translates to improved power quality for connected devices, ensuring reliable and efficient operation. Collectively, these visualizations not only demonstrate SMO's clear edge over PSO in optimizing buck converter performance but also emphasize the superior predictive capabilities of DNN models compared to RF models.

Fig.~\ref{fig13} illustrates the inductor current ripple. The dotted blue line represents the current achieved using PSO-optimized parameters, while the pink line corresponds to SMO-optimized parameters. A notable observation is that the SMO-optimized waveform exhibits smoother transitions and lower peak currents, resulting in a significant ripple reduction of approximately 15-20\%, as indicated by the annotation. This reduction in ripple reflects SMO's ability to produce a more consistent and controlled current flow, which is critical for efficient power delivery in buck converter applications.

\section{Conclusion \& Future Work}
This study introduces a data-driven DT framework for DC-DC buck converters. Using a low-power prototype and both empirical and synthetic datasets, the SMO+DNN approach shows clear advantages over traditional methods. Notably, SMO achieves the global optimum in 95\% of trials (vs.\ 65\% for PSO), requires 33\% fewer iterations, and results in 80\% fewer parameter constraint violations. The DNN model provides highly accurate predictions, with $R^2$ scores consistently above 0.998 for all key degradation parameters. Under accelerated thermal ageing, the framework reduces voltage ripple by 20--25\% and inductor current ripple by 15--20\%, ensuring more stable converter operation. These results demonstrate the framework’s effectiveness in early detection of component degradation. Future work will target real-time embedded deployment, integration with GaN-based semiconductor technologies, broader application to other converter types and stressors.

\section*{Acknowledgement}
This research was supported by the Myron Zucker Student-Faculty Grant Program of the IEEE Industry Applications Society. The authors gratefully acknowledge this support, which made the successful completion of this work possible.

\end{document}